\documentclass[twocolumn,showpacs,preprintnumbers]{revtex4}
\usepackage{amssymb}
\usepackage{amsmath}
\usepackage{graphicx}
\usepackage{dcolumn}
\usepackage{bm}

\tighten
\input{tcilatex}

\begin{document}

\title{Quantum beat phenomenon presence in coherent spin dynamics of spin-2 $%
^{87}$Rb atoms in a deep optical lattice}
\author{Hua-Jun Huang and Guang-Ming Zhang}
\affiliation{Department of Physics, Tsinghua University, Beijing 100084, China}
\date{\today }

\begin{abstract}
Motivated by the recent experimental work (A. Widera, \textit{et al}, Phys.
Rev. Lett. 95, 19045), we study the collisional spin dynamics of two spin-2 $%
^{87}$Rb atoms confined in a deep optical lattice. When the system is
initialized as $|0,0\rangle $, three different two-particle Zeeman states
are involved in the time evolution due to the conservation of magnetization.
For a large magnetic field $B>0.8$ Guass, the spin coherent dynamics reduces
to a Rabi-like oscillation between the states $|0,0\rangle $ and $%
|1,-1\rangle $. However, under a small magnetic field, a general three-level
coherent oscillation displays. In particular, around a critical magnetic
field $B_{c}\simeq 0.48$ Guass, the probability in the Zeeman states $%
|2,-2\rangle $ exhibits a novel quantum beat phenomenon, ready to be
confirmed in future experiments.
\end{abstract}

\pacs{03.75.Lm, 03.75.Mn, 34.50.-s}
\maketitle

Spinor Bose-Einstein condensates (BEC) in purely optical traps has opened a
new direction in study of confined dilute atomic gases \cite%
{ho,ohmi,law,pu,ciobanu,koashi,you}, and many fascinating phenomena
originating from the spin degrees of freedom have been observed \cite%
{stamper-kurn,barrett,schmaljohann,chang,kuwamoto,higbie}. Among them,
coherent spin-exchange dynamics induced by interatomic collisions has been
investigated in several recent experiments with BEC condensates in an
optical lattice \cite{schmaljohann,chang,kuwamoto,sengstock}, where coherent
control of the evolution with a magnetic field to apply different phase
shifts to the spin states has also been demonstrated \cite{nature-physics}.

More recently, coherent spin-mixing oscillations in a Mott insulating state
of hyperfine spin-2 $^{87}$Rb atoms in deep optical lattices have been
reported \cite{widera-prl} in a system of many isolated of a pair of atoms
on each lattice site. A weakly damped Rabi-like oscillation between
two-particle Zeeman states with equal magnetization displays in the large
magnetic field, the oscillation frequency and amplitudes are precisely
determined by the spin-exchange couplings and the second order Zeeman
shifts. However, when the external magnetic field is weak, three different
two-particle Zeeman states with equal magnetization are involved, and a
general feature of three-level coherent oscillations are expected among the
possible spin collisional processes.

In this paper, we present the detailed calculations of the coherent
evolutions of a pair of spin-2 $^{87}$Rb atoms in a deep optical trap. When
the system is initialized as $|0,0\rangle $, for a large magnetic field
typically $B>0.8G$, the spin coherent dynamics reduces to a Rabi-like
oscillation between the states $|0,0\rangle $ and $|1,-1\rangle $, which has
been observed in the excellent experiment \cite{widera-prl}. However, for a
weak magnetic field, a three-level coherent oscillation displays. Around a
critical magnetic field $B_{c}\approx 0.48G$\ for the typical parameters, it
is more interesting that the probability in the Zeeman state $|2,-2\rangle $
exhibits a novel quantum beat phenomenon, which is possible to be observed
in further experiments.

For a pair of spin-2 boson atoms in a deep optical trap, the general
interaction is given by%
\begin{equation}
V=g_{0}\mathbf{P}_{0}+g_{2}\mathbf{P}_{2}+g_{4}\mathbf{P}_{4},
\end{equation}%
where $g_{F}=4\pi \hslash ^{2}a_{F}/M\int \mathbf{dr}\left\vert \varphi
\right\vert ^{4}$, with $M$ the mass of the atom, $a_{F}$ the s-wave
scattering lengths in the total spin-F channel, $\mathbf{P}_{F}$ the
projection operator for a total hyperfine spin-F state with $\mathbf{P}_{0}+%
\mathbf{P}_{2}+\mathbf{P}_{4}=1$, and $\varphi $ is the spin-independent
spatial wave function of the ground state in the potential trap.

For a pair of atoms, one could use the relation $\mathbf{S}_{1}\mathbf{\cdot
S}_{2}=-6\mathbf{P}_{0}-3\mathbf{P}_{2}+4\mathbf{P}_{4}$ to rewrite the
interaction into
\begin{equation}
V=c_{0}+c_{1}\mathbf{S}_{1}\mathbf{\cdot S}_{2}+c_{2}\mathbf{P}_{0},
\end{equation}%
with $c_{0}=(4g_{2}+3g_{4})/7$, $c_{1}=(g_{4}-g_{2})/7$, $%
c_{2}=(7g_{0}-10g_{2}+3g_{4})/7$ and $\mathbf{P}_{0}=\mathbf{\Theta }%
^{\dagger }\Theta /10$ \cite{ciobanu,koashi}, where $\Theta ^{\dagger
}=2b_{2}^{\dagger }b_{-2}^{\dagger }-2b_{1}^{\dagger }b_{-1}^{\dagger
}+b_{0}^{\dagger }b_{0}^{\dagger }$ creates a pair of atoms in the spin
singlet when applied to the\ vacuum, and$\ b_{m}^{\dagger }$ the creation
operator. Since the total spin angular momentum and its z-component are
conserved, the interaction can be diagonalized with the eigenvalues $\overset%
{\_}{E}_{j=4}=10c_{1}$, $\overset{\_}{E}_{j=2}=3c_{1}$, and $\overset{\_}{E}%
_{j=0}=c_{2}$, where $j$ is the good quantum number of the total angular
momentum. The corresponding eigenstates are described by $\left\vert \phi
_{j,m}\right\rangle $. It should be noted that one could not directly obtain
the atom populations in these eigenstates experimentally. Instead, the
populations of atom pairs in Zeeman states $\left\vert
m_{1},m_{2}\right\rangle $ could be easily obtained by absorption imaging
after a few milliseconds of\ time-of-flight (TOF) with a magnetic gradient
field.

When the atom pair is initialized as the Zeeman state $\left\vert
0,0\right\rangle $, other two-particle Zeeman states in the subspace of $%
m_{1}+m_{2}=0$ are involved in the time evolutions due to the conservation
of the total spin z-component magnetization. Introducing the following
notation
\begin{eqnarray}
\left\vert 1,-1\right\rangle &=&(\left\vert m_{1}=1,m_{2}=-1\right\rangle
+\left\vert m_{1}=-1,m_{2}=1\right\rangle )/\sqrt{2},  \notag \\
\left\vert 2,-2\right\rangle &=&(\left\vert m_{1}=2,m_{2}=-2\right\rangle
+\left\vert m_{1}=-2,m_{2}=2\right\rangle )/\sqrt{2},  \notag \\
\left\vert 0,0\right\rangle &=&\left\vert m_{1}=0,m_{2}=0\right\rangle ,
\end{eqnarray}%
we can explicitly express the two-particle Zeeman states in terms of linear
combinations of the eigenstates $\left\vert \phi _{j,m}\right\rangle $.%
\begin{eqnarray}
\left\vert 0,0\right\rangle &=&\sqrt{\frac{18}{35}}\left\vert \phi
_{4,0}\right\rangle -\sqrt{\frac{2}{7}}\left\vert \phi _{2,0}\right\rangle +%
\sqrt{\frac{1}{5}}\left\vert \phi _{0,0}\right\rangle ,  \notag \\
\left\vert 1,-1\right\rangle &=&\sqrt{\frac{16}{35}}\left\vert \phi
_{4,0}\right\rangle +\sqrt{\frac{1}{7}}\left\vert \phi _{2,0}\right\rangle -%
\sqrt{\frac{2}{5}}\left\vert \phi _{0,0}\right\rangle ,  \notag \\
\left\vert 2,-2\right\rangle &=&\sqrt{\frac{1}{35}}\left\vert \phi
_{4,0}\right\rangle +\sqrt{\frac{4}{7}}\left\vert \phi _{2,0}\right\rangle +%
\sqrt{\frac{2}{5}}\left\vert \phi _{0,0}\right\rangle ,
\end{eqnarray}%
where the coefficients in front of the eigenstates $\left\vert \phi
_{j,m}\right\rangle $ for each Zeeman states $\left\vert
m_{1},m_{2}\right\rangle $ are just given by the Clebsh-Gordon (C-G)
coefficients. Therefore, the wave function of the system at any time $t$ can
be expressed
\begin{equation*}
\left\vert \psi _{t}\right\rangle =C_{0}(t)\left\vert 0,0\right\rangle
+C_{1}(t)\left\vert 1,-1\right\rangle +C_{2}(t)\left\vert 2,-2\right\rangle ,
\end{equation*}%
with
\begin{widetext}
\begin{eqnarray}
C_{0}(t) &=&\frac{18}{35}\exp (-i\overset{\_}{E}_{4,0}t)+\frac{2}{7}\exp (-i%
\overset{\_}{E}_{2,0}t)+\frac{1}{5}\exp (-i\overset{\_}{E}_{0,0}t),  \notag
\\
C_{1}(t) &=&\frac{\sqrt{288}}{35}\exp (-i\overset{\_}{E}_{4,0}t)-\frac{\sqrt{%
2}}{7}\exp (-i\overset{\_}{E}_{2,0}t)-\frac{\sqrt{2}}{5}\exp (-i\overset{\_}{%
E}_{0,0}t),  \notag \\
C_{2}(t) &=&\frac{\sqrt{18}}{35}\exp (-i\overset{\_}{E}_{4,0}t)+\frac{\sqrt{8%
}}{7}\exp (-i\overset{\_}{E}_{2,0}t)+\frac{\sqrt{2}}{5}\exp (-i\overset{\_}{E%
}_{0,0}t).
\end{eqnarray}
\end{widetext}The corresponding probabilities in each Zeeman state are
easily derived
\begin{eqnarray}
P_{0} &=&\frac{473}{1225}+\frac{72}{245}\cos \omega _{1}t+\frac{4}{35}\cos
\omega _{2}t+\frac{36}{175}\cos \omega _{3}t,  \notag \\
P_{1} &=&%
\frac{436}{1225}%
-\frac{48}{245}\cos \omega _{1}t+\frac{4}{35}\cos \omega _{2}t-\frac{48}{175}%
\cos \omega _{3}t,  \notag \\
P_{2} &=&%
\frac{316}{1225}%
-\frac{24}{245}\cos \omega _{1}t-\frac{8}{35}\cos \omega _{2}t+\frac{12}{175}%
\cos \omega _{3}t,  \notag
\end{eqnarray}%
where $\omega _{1}=\overset{\_}{E}_{4,0}-\overset{\_}{E}_{2,0}$, $\omega
_{2}=\overset{\_}{E}_{2,0}-\overset{\_}{E}_{0,0}$, and $\omega _{3}=\overset{%
\_}{E}_{4,0}-\overset{\_}{E}_{0,0}$. Clearly, the spin dynamics of
the system displays a three-level coherent oscillation with three
different frequencies and comparable amplitudes. We have plotted
these probabilities in Fig.1. We would like to emphasize that the
oscillation frequencies are determined by the eigenvalues, instead
of the off-diagonal matrix elements, of the $3\times 3$ model
Hamiltonian matrix.
\begin{figure}[tbp]
\begin{center}
\includegraphics[width=2.0in]{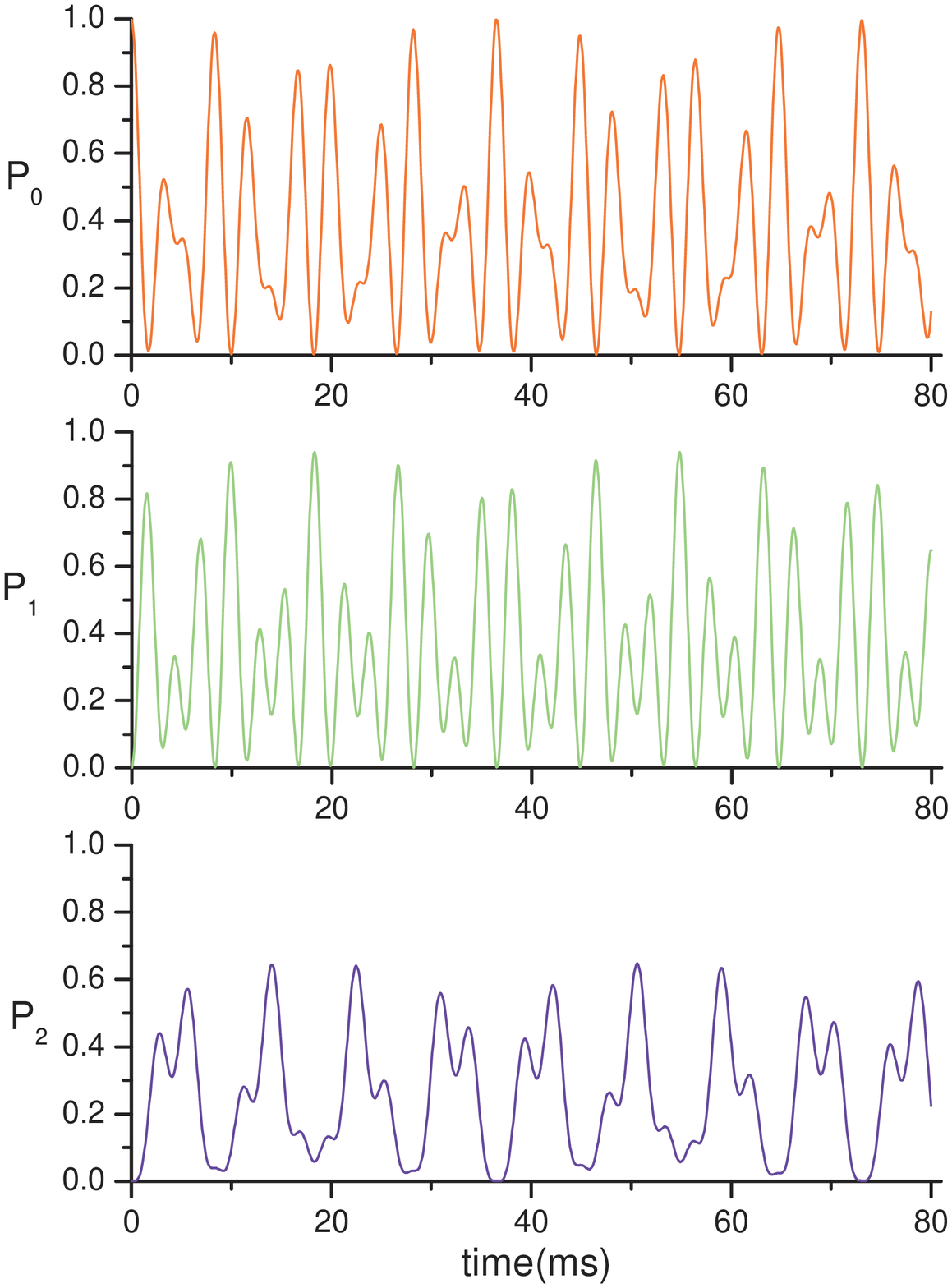}
\end{center}
\caption{Probabilities of the two-particle Zeeman states in the absence of
magnetic field.}
\end{figure}

In the presence of a magnetic field, the linear and quadratic Zeeman terms
should be included into the interaction Hamiltonian \cite{sengstock}
\begin{equation}
H=V-p(S_{1z}+S_{2z})+q(S_{1z}^{2}+S_{2z}^{2}),
\end{equation}%
where $p=\mu _{B}B/2\hbar $, the ground state hyperfine splitting $\omega
_{hfs}\approx 2\pi \times 6.835GHz$ for $^{87}$Rb atoms, and $q=p^{2}/\omega
_{hfs}\approx 450B^{2}Hz/G^{2}$. In this case, the total spin angular
momentum is \textit{no longer} conserved, but its z-component is still
conserved. Therefore, we can work in the subspace of $m_{1}+m_{2}=0$, in
which the second order Zeeman shift can influence the spin dynamics. In the
complete bases of $\left\vert 0,0\right\rangle $, $\left\vert
1,-1\right\rangle $, and $\left\vert 2,-2\right\rangle $, the effective
Hamiltonian could be expressed as:%
\begin{equation}
H_{eff}=\left[
\begin{array}{ccc}
9.82J & 7.02J & -0.05J \\
7.02J & 8.15J+2q & 3.35J \\
-0.05J & 3.35J & 3.22J+8q%
\end{array}%
\right] ,
\end{equation}%
where we have used $a_{0}=89.4a_{B}$, $a_{2}=94.5a_{B}$, and $%
a_{4}=106.0a_{B}$ for $^{87}$Rb atoms \cite{ciobanu}, and $J=\frac{4\pi
\hbar ^{2}n^{2}}{M}\int \mathbf{dr}\left\vert \varphi \right\vert ^{4}$ is
typically chosen as $129.02Hz$. To solve this Hamiltonian, one has to employ
the numerical method. Using the harmonic wave function to approximate $%
\varphi $, one could estimate%
\begin{eqnarray}
Ja_{B} &\sim &\frac{a_{B}\hbar ^{2}}{\pi M}\sqrt{\frac{2}{\pi }}(\frac{k}{%
\hbar }\sqrt{2V_{0}M})^{3/2}  \notag \\
&=&4.08v^{3/4}Hz,
\end{eqnarray}%
where $V_{0}=v\frac{\hbar ^{2}k^{2}}{2M}$ is the depth of the lattice
potential. In the Mott insulating regime, $v$ is large and we simply choose $%
v=100$. In Fig.2a, the eigenvalues of the effective Hamiltonian ($\overset{\_%
}{E}_{1}>\overset{\_}{E}_{2}>\overset{\_}{E}_{3}$) have been presented as a
function of the magnetic field.
\begin{figure}[tbp]
\begin{center}
\includegraphics[width=2.0in]{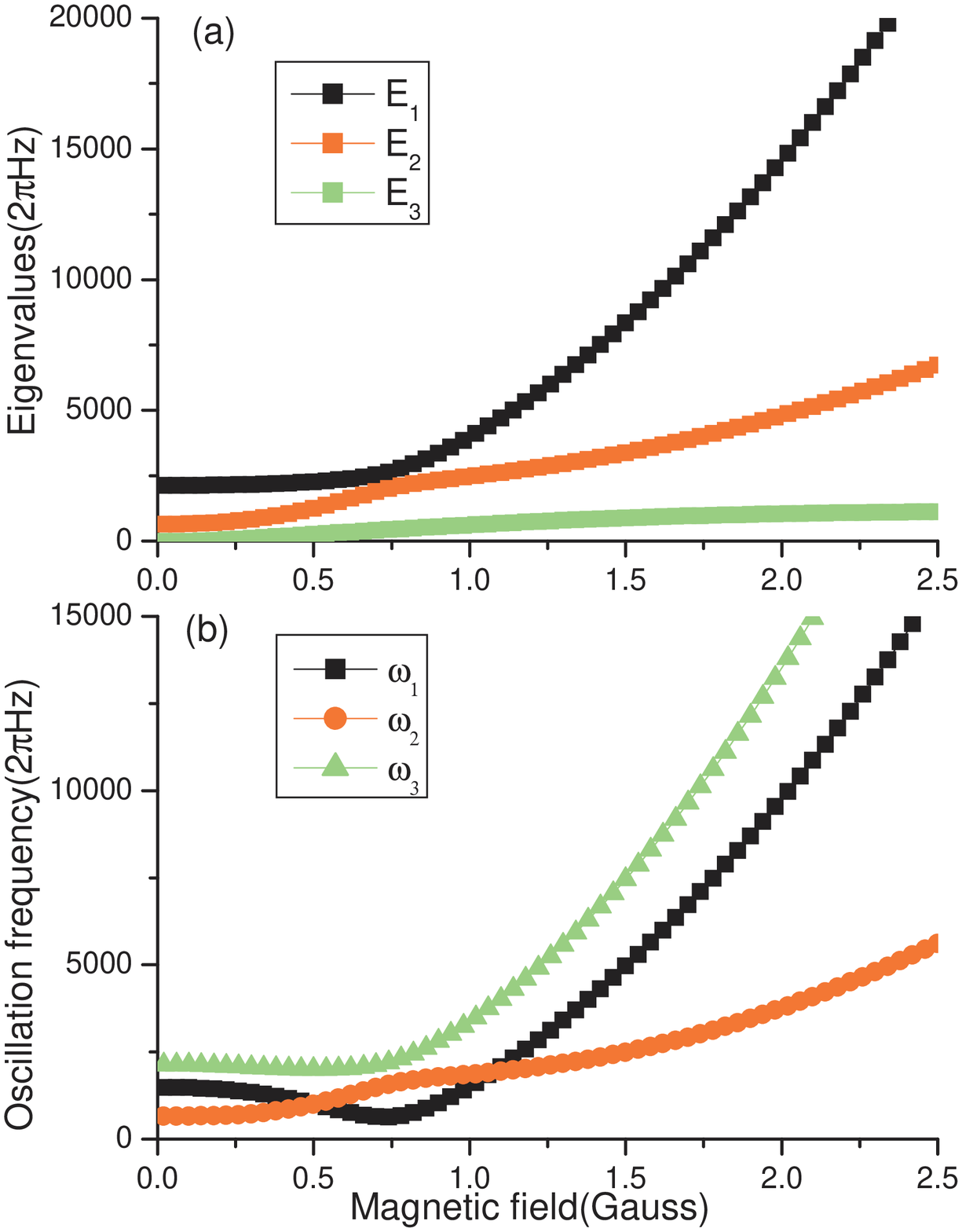}
\end{center}
\caption{Eigenvalues of the effective model in (a) and the corresponding
three oscillation frequencies in (b) as a function of magnetic field.}
\end{figure}

In the presence of the second order Zeeman shift, three two-particle Zeeman
states can still be expressed in terms of three eigenstates%
\begin{eqnarray}
\left\vert 0,0\right\rangle &=&U_{11}\left\vert \overset{\_}{E}%
_{1}\right\rangle +U_{12}\left\vert \overset{\_}{E}_{2}\right\rangle
+U_{13}\left\vert \overset{\_}{E}_{3}\right\rangle ,  \notag \\
\left\vert 1,-1\right\rangle &=&U_{21}\left\vert \overset{\_}{E}%
_{1}\right\rangle +U_{22}\left\vert \overset{\_}{E}_{2}\right\rangle
+U_{23}\left\vert \overset{\_}{E}_{3}\right\rangle ,  \notag \\
\left\vert 2,-2\right\rangle &=&U_{31}\left\vert \overset{\_}{E}%
_{1}\right\rangle +U_{32}\left\vert \overset{\_}{E}_{2}\right\rangle
+U_{33}\left\vert \overset{\_}{E}_{3}\right\rangle ,
\end{eqnarray}%
where nine coefficients are calculated numerically and are plotted as a
function of the magnetic field in Fig.3a. It can be seen that $U_{11}$ and $%
U_{33}\,$become very small and reduce quickly to zero as the magnetic field
is increased. When $U_{11}=0$, the initial state is just a superposition of
two eigenstates only $\left\vert 0,0\right\rangle \simeq U_{12}\left\vert
\overset{\_}{E}_{2}\right\rangle +U_{13}\left\vert \overset{\_}{E}%
_{3}\right\rangle $, so it will not jump into third eigenstate $\left\vert
\overset{\_}{E}_{1}\right\rangle $. Similar to the previous procedure, the
probabilities of the Zeeman states are obtained as the form of
\begin{eqnarray}
P_{0} &=&A_{00}+A_{01}\cos \omega _{1}t+A_{02}\cos \omega _{2}t+A_{03}\cos
\omega _{3}t,  \notag \\
P_{1} &=&A_{10}+A_{11}\cos \omega _{1}t+A_{12}\cos \omega _{2}t+A_{13}\cos
\omega _{3}t,  \notag \\
P_{2} &=&A_{20}+A_{21}\cos \omega _{1}t+A_{22}\cos \omega _{2}t+A_{23}\cos
\omega _{3}t,  \notag
\end{eqnarray}%
with the amplitudes as combinations of the coefficients $U_{ij}$
\begin{widetext}
\begin{eqnarray}
A_{00}
&=&U_{11}^{4}+U_{12}^{4}+U_{13}^{4},%
\,A_{10}=U_{11}^{2}U_{12}^{2}+U_{12}^{2}U_{22}^{2}+U_{13}^{2}U_{23}^{2},%
\,A_{20}=U_{11}^{2}U_{31}^{2}+U_{12}^{2}U_{32}^{2}+U_{13}^{2}U_{33}^{2},%
\text{\ \ \ \ }  \notag \\
A_{01} &=&2U_{11}^{2}U_{12}^{2},\text{ \ \ \ \ \ \ \ \thinspace \thinspace }%
A_{11}=2U_{11}U_{12}U_{21}U_{22},\text{\ \ \ \ \ \ \ \ \ \ \ \ }%
A_{21}=2U_{11}U_{12}U_{31}U_{32},  \notag \\
\text{ }A_{02} &=&2U_{12}^{2}U_{13}^{2},\,\ \ \ \ \ \ \ \
\,A_{12}=2U_{12}U_{13}U_{22}U_{23},\ \ \ \ \ \ \ \ \ \ \ \
A_{22}=2U_{12}U_{13}U_{32}U_{33},\text{ \ \ }  \notag \\
\text{\ }A_{03} &=&2U_{11}^{2}U_{13}^{2},\text{ \ \ \ \ \ \ \ \thinspace
\thinspace }A_{13}=2U_{11}U_{13}U_{21}U_{23},\ \ \ \ \ \ \ \ \ \ \ \
A_{23}=2U_{11}U_{13}U_{31}U_{33},
\end{eqnarray}
\end{widetext}and the frequencies are given by
\begin{equation*}
\omega _{1}=\overset{\_}{E}_{1}-\overset{\_}{E}_{2},\omega _{2}=\overset{\_}{%
E}_{2}-\overset{\_}{E}_{3},\omega _{3}=\overset{\_}{E}_{1}-\overset{\_}{E}%
_{3}.
\end{equation*}%
There are simple relations between the amplitudes of each oscillations and
the frequencies:%
\begin{equation}
A_{0n}+A_{1n}+A_{2n}=\delta _{n,0},\text{ \ }\omega _{3}=\omega _{1}+\omega
_{2}.
\end{equation}%
\begin{figure}[tbp]
\begin{center}
\includegraphics[width=3.0in]{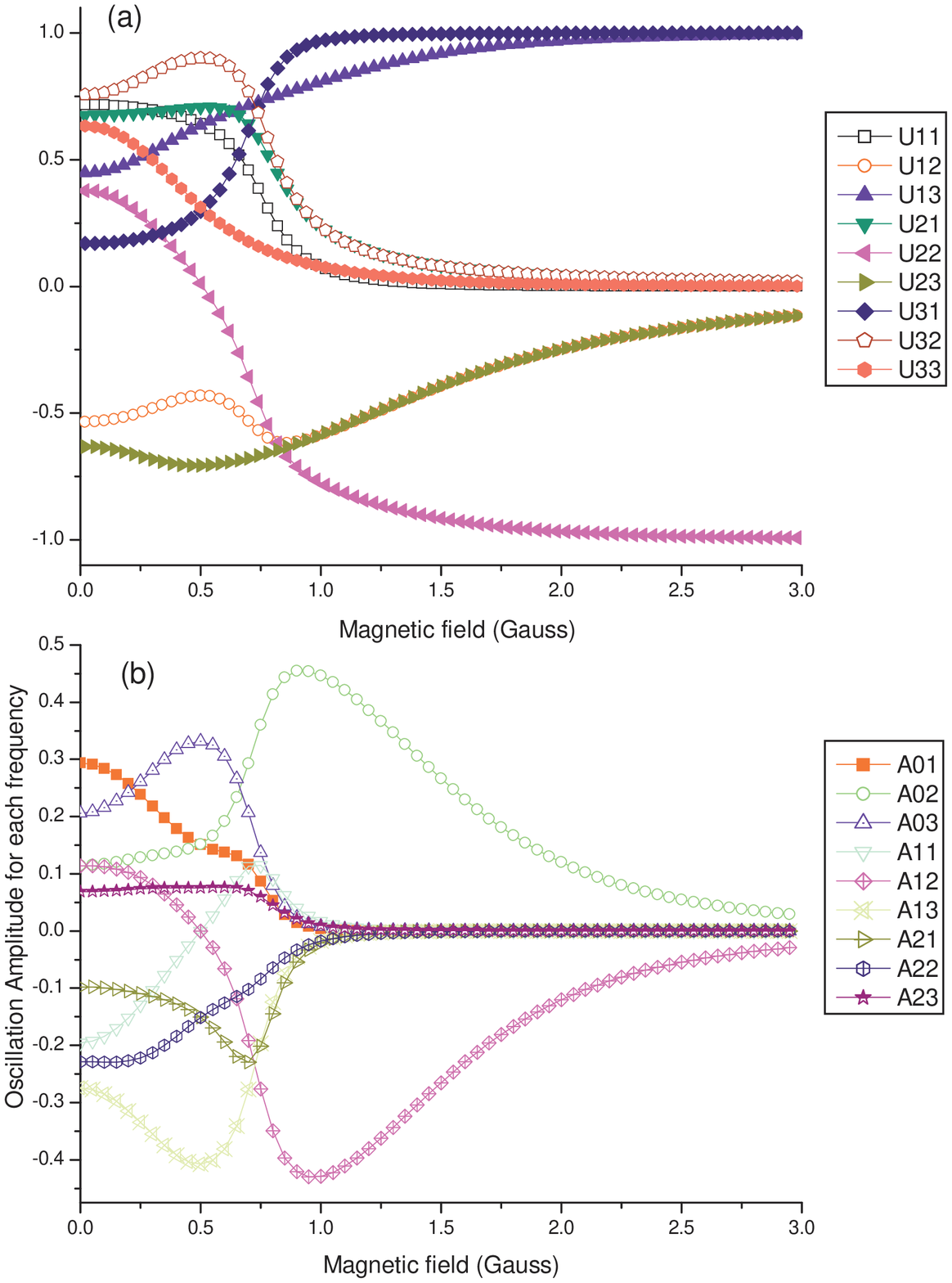}
\end{center}
\caption{Expansion coefficients of the two-particle Zeeman states in terms
of eigenstates $U_{ij}$ in (a) and the amplitudes of each frequency
oscillation $A_{ij}$ in (b) as a function of magnetic field.}
\end{figure}

As the magnetic field $B$ gradually increases, the frequencies $\omega _{i}$
of the spin oscillations are determined by the eigenvalues $\overset{\_}{E}%
_{n}$, while the amplitudes of each frequency oscillation are determined by
the coefficients $U_{ij}$. By focusing on the amplitudes $A_{mn}$ for each
oscillation frequency as the functions of the magnetic field shown in
Fig.3b, we can clearly demonstrate how the system crossovers from a
three-level ($\left\vert 2,-2\right\rangle $, $\left\vert 1,-1\right\rangle $%
, $\left\vert 0,0\right\rangle $) coherent oscillation without any selection
rules\ into a two-level ($\left\vert 1,-1\right\rangle $, $\left\vert
0,0\right\rangle $) Rabi-like oscillation.

When the large magnetic field is typically larger than $0.8G$, which is a
reasonable magnitude compared to the experimental parameters \cite%
{widera-prl}, all amplitudes $A_{mn}$ begin to\ fall quickly down to zero
except for\emph{\ }$A_{02}$\emph{\ }and\emph{\ }$A_{12}$. Thus, $P_{2}(t)$
could be neglected and $P_{0}$ and $P_{1}$ are reduced into an oscillation
with a cosine form, i.e., the system oscillates between $\left\vert
0,0\right\rangle $ and $\left\vert 1,-1\right\rangle $ like a Rabi
oscillation. In Fig.4, we have plotted the probabilities of the three Zeeman
states at $B=1.2G$. To find out the more detailed reason, one must look into
the coefficients $U_{ij}$ shown in Fig.3b: $U_{11}$ and $U_{33}\,$become
very small and reduce quickly to zero, $A_{m1}$ and $A_{m3}$ ($m=1,2,3)$
including $U_{11}$ fall down to zero, and $A_{22}$ including $U_{33}$ falls
down to zero too. In the extreme limit of $q/J\gg 1$, all $U_{ij}$ are zero
except $U_{13},\,U_{22,}\,U_{31}$ equal to $1$. Then the initial state $%
\left\vert 0,0\right\rangle $ just corresponds to one eigenstate, so there
will be no spin-exchange collisional dynamics.
\begin{figure}[tbp]
\begin{center}
\includegraphics[width=2.0in]{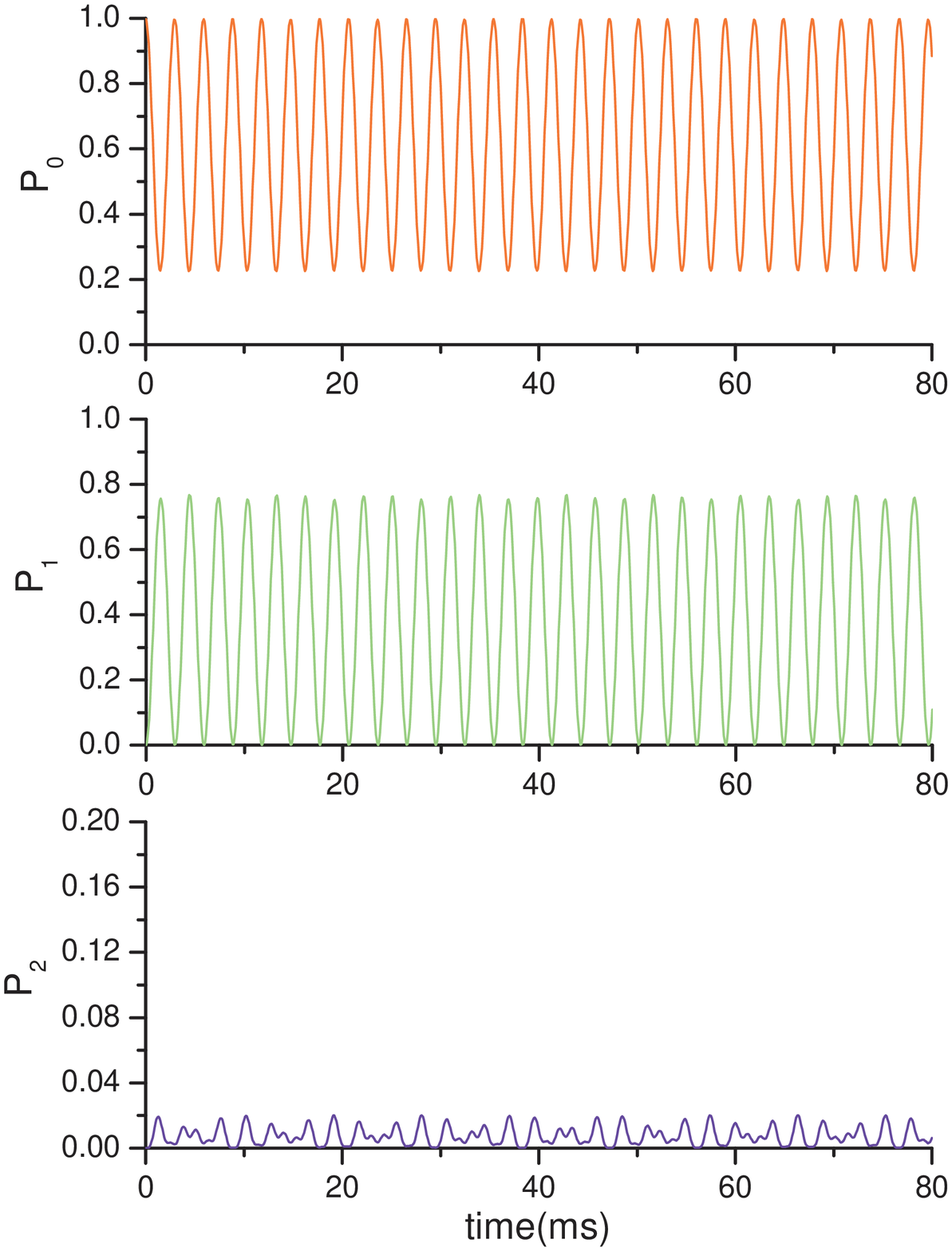}
\end{center}
\caption{Probabilities of the two-particle Zeeman states in the large
magnetic field $B=1.2G$.}
\end{figure}

However, in the weak magnetic field regime, the amplitudes of three
frequencies in $P_{0}$, $P_{1}$ and $P_{2}$ are all finite and comparable,
displaying the three-level coherent oscillations. As $B$ is gradually
increasing, the different eigenvalues grow with different slopes. Around a
critical magnetic field $B_{c}\approx 0.48G$,$\,$\ $\omega _{1}$ and $\omega
_{2}$ are very close to\emph{\ }each other,\emph{\ }$A_{01}\simeq
A_{02}\simeq 0.15$, $\,A_{03}\simeq 0.35$, $\ A_{11}\simeq A_{12}\simeq 0$, $%
\,A_{13}\simeq -0.42$,\emph{\ }$\,A_{21}\simeq A_{22}\simeq -0.15$,$\ $and$\
A_{23}\simeq 0.07$.\emph{\ }Since $A_{23}$\emph{\ }is small with compared to
the beat amplitude (roughly $2A_{21}$\ or $2A_{22}$), there appears a clear
quantum beating phenomenon presence in the probability of the Zeeman state $%
|2,-2\rangle $. Meanwhile, the corresponding probability $P_{0}(t)$\
exhibits a combination of a beating\ oscillation and a sinusoidal
one with the frequency $2\omega $, because the sinusoidal amplitude
$A_{03}\ $is comparable to the beat amplitude (roughly $2A_{01}$\ or
$2A_{02}$). $P_{1}(t) $ just corresponds to a single frequency
oscillation. In Fig.5, we have plotted the three probabilities at
$B=0.47G$. Considering that this novel phenomenon is very sensitive
to the external magnetic field, the critical point\ might be
accurately determined by fine-tuning the external field
experimentally.
\begin{figure}[tbp]
\begin{center}
\includegraphics[width=2.0in]{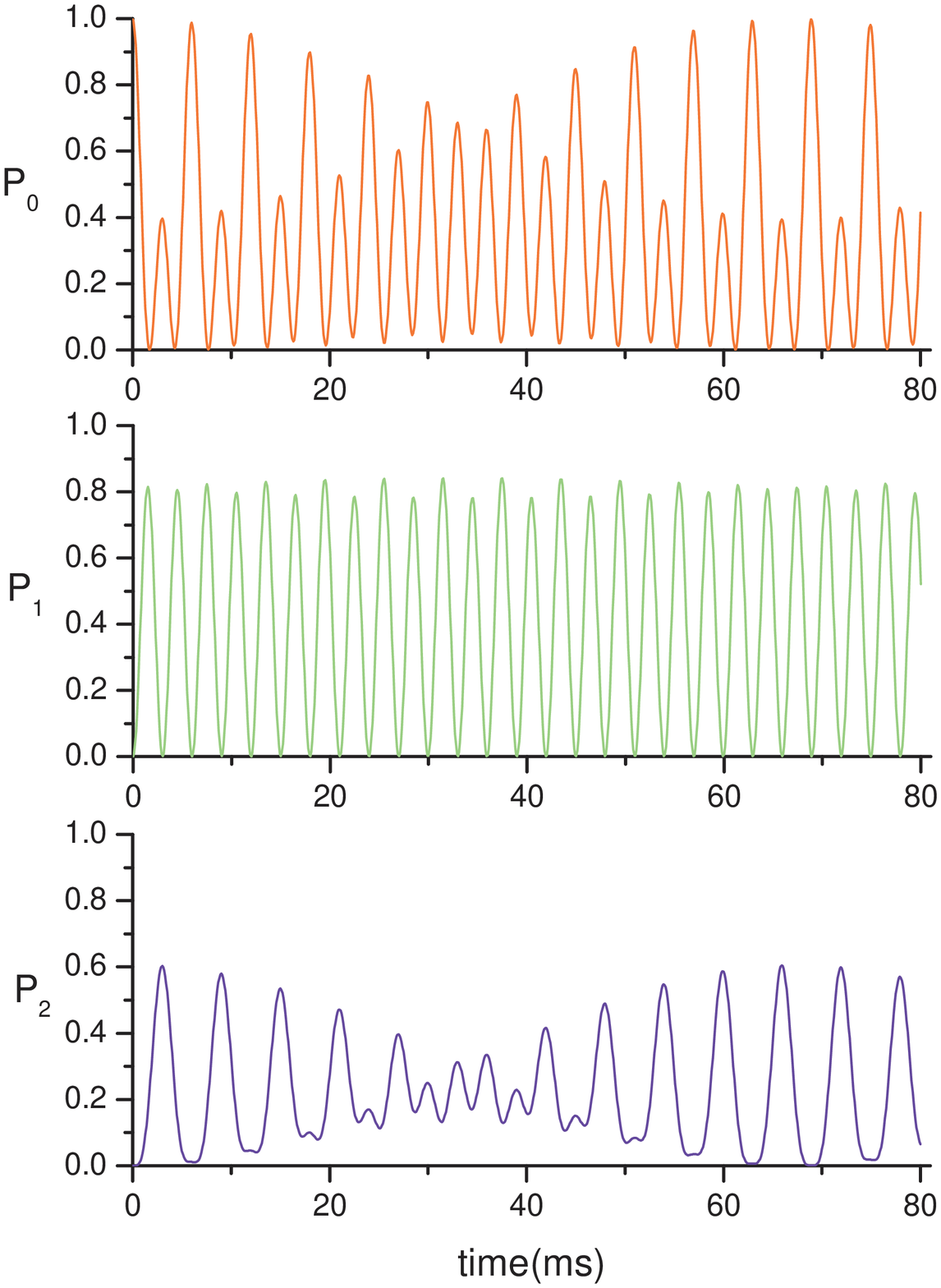}
\end{center}
\caption{Probabilities of the two-particle Zeeman states at
$B=0.47G$ around the critical magnetic field.}
\end{figure}

In conclusion, we have studied the collisional spin dynamics of an isolated
spin-2 $^{87}$Rb atom pairs confined in a deep optical lattice. Although our
calculations could not consider the damping of the oscillations, some clear
experimental predictions are given in the weak magnetic field limit, and we
have also demonstrated the presence of a crossover from the three-level to
two-level coherent spin dynamics. In particular, when the system is
initialized as $|0,0\rangle $ and the magnetic field $B_{c}\approx 0.48G$,
the probability in the two-particle Zeeman state $|2,-2\rangle $ will
exhibit a quantum beat phenomenon, which is ready to be confirmed in future
experiments.

The authors are indebted to Prof. Li You for his many stimulating
discussions, and G. M. Zhang is supported by NSF-China (Grant No. 10125418
and 10474051).

\end{document}